\documentclass[aps,preprintnumbers,prl,twocolumn,superscriptaddress,nofootinbib]{revtex4}

\usepackage{epsfig,latexsym,cancel,amssymb,amsmath}
\usepackage{graphicx}
\usepackage{epstopdf}
\usepackage{bbold}
\usepackage{numprint}

\renewcommand{\sec}{\ensuremath{\mathrm{s}}}

\newcommand{\kg}{\ensuremath{\mathrm{kg}}}

\newcommand{\fm}{\ensuremath{\mathrm{fm}}}
\newcommand{\cm}{\ensuremath{\mathrm{cm}}}
\newcommand{\km}{\ensuremath{\mathrm{km}}}

\newcommand{\eV}{\ensuremath{\mathrm{eV}}}
\newcommand{\keV}{\ensuremath{\mathrm{keV}}}
\newcommand{\MeV}{\ensuremath{\mathrm{MeV}}}
\newcommand{\GeV}{\ensuremath{\mathrm{GeV}}}

\newcommand{\bet}{\ensuremath{{}^8\mathrm{B}}}

\newcommand{\keVee}{\ensuremath {\mathrm{keVee}}}

\newcommand{\Er}{\ensuremath {E_R}}
\newcommand{\Ev}{\ensuremath {E_\mathrm{v}}}
\newcommand{\cpd}{\ensuremath {\mathrm{cpd}}}
\newcommand{\days}{\ensuremath {\mathrm{days}}}

\newcommand{\nub}{\ensuremath {\nu_{b}}}

\newcommand{\Neff}{\ensuremath {\mathcal{N}_{\mathrm{eff}}}}

\newcommand{\Enu}{\ensuremath {E_\nu}}

\begin{document}

\title{ Dark Matter or Neutrino recoil? 
Interpretation of Recent Experimental Results. }

\author{Maxim Pospelov}
\affiliation{Perimeter Institute, Waterloo, Ontario N2L 2Y5, Canada}
\affiliation{Department of Physics and Astronomy, University of Victoria, Victoria, BC, V8P 5C2, Canada}

\author{Josef Pradler}
\affiliation{Department of Physics and Astronomy, Johns Hopkins University, Baltimore, MD 21210, USA}

\begin{abstract}
  The elastic nuclear recoil signal, being under intense scrutiny by
  multiple underground experiments, can be interpreted either as
  coming from the interaction of nuclei with WIMP dark matter or from
  the scattering of new species of MeV-energy neutrinos.  The most
  promising model for the latter case is a neutrino $\nu_b$ that
  interacts with baryon number, and with a flux sourced by the
  oscillations of regular solar $^8$B neutrinos. We re-analyze this
  model in light of the latest experimental results. In contrast to
  the light-DM interpretation of various tentative positive signals
  (anomalies) that is now seriously challenged by the negative results
  of the LUX experiment, the neutrino interpretation remains a viable
  explanation to most of the anomalies. Considering future prospects,
  we show that the superCDMS experiment alone, when equipped with Ge
  \textit{and} Si detectors, will be able to detect $\nu_b$ and
  discriminate the model from a light DM interpretation. In addition,
  we also provide the forecast for the new CRESST-II run that now
  operates with new detectors and diminished backgrounds.
\end{abstract}

\maketitle

\section{Introduction}
\label{sec:introduction}

The search for a non-gravitational detection of dark matter (DM) has
entered a new phase with more and more experiments reporting their
results at ever increasing limits of sensitivity.  Indeed, if DM is
composed of weakly interacting massive particles (WIMPs), there is a
real chance that occasional events of WIMP-atom scattering may be
recorded in low-background experiments~\cite{PhysRevD.31.3059}.  The
weakness of DM-nucleus scattering can only be compared to neutrino
detection, which is also very challenging on the account of a
neutrino's weak interaction with ordinary matter. Necessarily then,
sensitive neutrino and WIMP detectors usually consist of fairly large
target mass, and due to the simplicity of the underlying event ({\em
  e.g.}~nuclear recoil) register only few channels. This stands in
sharp contrast with particle physics experiments at colliders, where,
given enough complexity and sophistication of detectors, hundreds and
thousands of channels can be registered as part of one event. Given
the rather simple form of WIMP scattering events, even in the case of
a positive signal detection, there will be a serious challenge in
distinguishing it from other potential origins of nuclear recoil.  One
idea that has been put forward recently
\cite{Pospelov:2011ha,Harnik:2012ni,Pospelov:2012gm} shows that the
current generation of WIMP detectors is also sensitive to the
combination of non-standard neutrino oscillation and interactions,
especially in the regime of recoil similar to that created by
relatively light, $10$ GeV and less, WIMP particles.  Moreover, it has
been suggested that some \cite{Pospelov:2011ha,Pospelov:2012gm} (if
not all) of the existing hints on a positive signal can be explained
within the model of an extra neutrino species that interacts with the
baryonic current.

The purpose of this paper is to revisit this idea, and to give an
update to the interpertation of a variety of direct DM detection
experimental results in terms of the recoil of a new neutrino species
that originates from ordinary solar neutrinos due to oscillation.
Since our last publication on the subject~\cite{Pospelov:2012gm},
there has been a series of developments in the field, notably a new
excess of events reported by the CDMS-II collaboration for their
silicon-made detectors~\cite{Agnese:2013rvf}, and a negative result
from the LUX experiment~\cite{Akerib:2013tjd} that strongly disfavors
the interpretation of many anomalies as a WIMP DM
signal~\cite{Gresham:2013mua,DelNobile:2013gba}. Given the immense amount of both
experimental and theoretical interest to the subject, and especially
to the light WIMP interpretations of the direct detection anomalies,
it is important to understand the status of alternative explanations
for the signal.

This paper is organized as follows: In the next two sections we give
an overview of the neutrino oscillation portal idea, and introduce the
model for a new neutrino neutrino $\nu_b$ that has stronger-than-weak
interaction strength $G_B$ with the baryonic current. After that we
adress new experimental results, one by one, discussing implications
of the liquid xenon
experiments~\cite{Angle:2011th,Aprile:2012nq,Akerib:2013tjd}, CDMS-II
silicon~\cite{Agnese:2013rvf}, and
CoGeNT~\cite{Aalseth:2011wp,Aalseth:2012if}.  The next-to-last section
discusses details of how the recoil of $\nu_b$ states and light dark
matter can be distniguished in practice, given more statistic planned
for the new run of CRESST-II\cite{Angloher:2011uu}, and making
projections for superCDMS~\cite{Akerib:2006rr}.

\section{Neutrino Oscillation Portal}
\label{sec:oscillation}

Considering a light fourth neutrino, we have two limiting cases what
regards the mass splitting relative to the three, mostly active
flavors. 
\begin{enumerate}
\item Sizable mass squared differences $|\Delta m_{4i}^2|\gg |\Delta
  m_{ij}^2|$ for $i,j =1,2,3.$ This scenario is typically considered
  in the context of the short baseline anomalies
  LSND~\cite{Aguilar:2001ty}, MiniBooNE~\cite{AguilarArevalo:2010wv},
  and the deficit of reactor neutrinos; see \cite{Kopp:2013vaa} and
  references therein. In the presence of enhanced interactions, this
  scenario is difficult to reconcile with observations: matter effects
  will likely have a substantial impact on the flavor evolution, and
  enhanced values of a new interaction will be tightly constrained by
  the total counting rate of neutrino-related events.
\item A new state that is (nearly) degenerate with one of the SM
  massive neutrinos $n_i$, $\Delta m^2 \equiv |\Delta m_{4i}^2| \ll
  |\Delta m_{ji}^2|$ for $j\neq i,4$. In this case the
  classification into short- and long-baseline neutrino experiments
  will remain largely intact; $\Delta m^2$ becomes the controlling
  parameter of flux in the new flavor $\nu_b$ generated from
  oscillation. We will show that in the limit of $\Delta m^2\to 0$ the
  flux in the fourth state that would be generated from oscillation
  can vanish altogether.  Therefore, there is always a choice of
  $\Delta m^2$ small enough that there is no impact on any of the
  neutrino oscillation experiments other than via pair-production of
  $\nu_b$.
\end{enumerate}
We shall pursue option (2) in the following and call it the ``neutrino
oscillation portal.''

We start from the general expression of flavor conversion
probabilities $ P_{\beta \alpha}(t) \equiv |\langle \nu_{\beta} |
\nu_{\alpha}(t) \rangle|^2 $ for $N$ neutrinos $| \nu_{\alpha} \rangle
$ that related to their mass eigenstates $|n_i\rangle$ via the unitary
transformation $ | \nu_{\alpha} \rangle = \sum_{i=1}^N U_{\alpha
  i}^{*} |n_i\rangle$.
Let us consider a hierarchy of masses between two groups $A= \left\{
  n_1 ,\dots , n_{N_A} \right\}$ and $ B = \left\{ n_{N_A+1} ,\dots
  ,n_{N} \right\}$ and assume baselines such that phases among
respective group members are negligible. Then, there is only one
common dominating phase, namely, between members of $A$ and $B$,
$\phi_{AB} \simeq \Delta m_{AB}^2 L / (2\Enu)$; $L\simeq c t$ is the
propagation distance and $\Enu$ is the relativistic neutrino
energy. The appearance probability is then given by,
\begin{align}
  P_{\beta \alpha} & = \left| \sum_k U_{\alpha k}^{*} U_{\beta k}
    \exp{\left( -i \frac{\Delta m_{k1}^2 L}{2\Enu} \right)} \right|^2
  \nonumber \\ & = \sin^2 \left( 2\theta_{\beta\alpha}^{\rm eff}
  \right) \sin^2 \left( \frac{\Delta m_{AB}^2 L}{4\Enu} \right) \quad
  (\alpha\neq \beta),
\end{align}
where the second relation is written in analogy to the two-flavor
case. The sine-squared of the effective angle reads (see,
\textit{e.g.}~\cite{Giunti:2007ry}),
\begin{align}
\label{eq:sine-eff}
  \sin^2 \left( 2\theta_{\beta\alpha}^{\rm eff} \right) = 4 \left|
    \sum_{k>N_A} U^{*}_{\alpha k } U_{\beta k} \right|^2 . 
\end{align} 
One can replace $ \sum_{k>N_A} $ with $ \sum_{k<N_A+1} $ because of
unitarity of $U$.

Oscillation from $\alpha= e, \mu, \tau$ into a \textit{new} state
$\beta = b$ is protected if the transition amplitude vanishes to
desired accuracy, $\sin^2 ( 2\theta_{b\alpha}^{\rm eff}) = 0$. This
can be achieved as follows:
Consider, for concreteness, the case of four light neutrinos
$n_{1,\dots 4}$ and adopt the notation in which $n_{1,2,3}$ are mostly
SM states, with PMNS-like mixing matrix $V_3$, and in which $n_4$ is
mostly $\nu_b$.
Then, the unitary $4\times 4$ mixing matrix $U$ can be defined as a
product, such, that $V_3$ stands to the left of new rotations $R_{i4}$
in the $i4$ plane by angles $\theta_{i4}$,
\begin{align}
\label{eq:maximsU}
  U = V_3  \prod_{i=1}^3 R_{i4}(\theta_{i4}).
\end{align}
If all new angles $\theta_{i4}$ between the groups $A$ and $B$ for
which a non-negligible phase $\phi_{AB}$ exists are zero, then,
indeed, $\sin^2 ( 2\theta_{b\alpha}^{\rm eff}) = 0$. This gives rise
to the notion of what we shall call the ``neutrino oscillation portal.''

In the following, we shall consider the case $A=\{n_1, n_2,n_4 \}$ and
$B=\{n_3\}$ and assume a fine mass splitting $\Delta m^2 <
10^{-8}\,\eV^2$ between $n_2$ and $n_4$. Keeping only one mixing angle
$\theta_b = \theta_{24}\neq 0$, $ \theta_{14},\theta_{34} =0$ the
fourth neutrino $n^0_4$ is mixed into $n^0_2$ as a small perturbation,
such that $n_{2,4}$ are the true propagating fields,%
\footnote{The relation between fields is $\vec \nu = U \vec n $ and we
  have used the notation $ \vec n^0 = V_3^{\dag} \vec \nu $ where we
  imagine $V_3$ quadratic and diagonal in the fourth state.}
\begin{align*}
  n_2 = \cos\theta_b n_2^0 - \sin\theta_b n_4^0 , \quad n_4 =
   \sin\theta_b n_2^0 +   \cos\theta_b n_4^0  .
\end{align*}
In this scenario, one finds of the vacuum transition probabilities
from flavor $\alpha = e,\mu,\tau$ into $\nu_b$
\begin{align}
  P_{b\alpha} < 10^{-4}\quad \text{for} \quad L\leq 10^3\,\km,
\end{align}
which holds for $E \geq 1\,\MeV$ and any value of $\theta_b$.%
\footnote{We have used SM angles $\sin^2(2\theta_{12})= 0.86$,
  $\sin^2(2\theta_{23})\simeq 1$, $\sin^2(2\theta_{13}) = 0.09$ in the
  standard parameterization of $V_3$ and $|\Delta m_{31}^2 |\simeq
  2.3(1) \times 10^{-3}\, \eV^2 $ and $\Delta m_{\odot}^2 = \Delta
  m_{21}^2 \simeq 7.5(2) \times 10^{-5}\, \eV^2$ for atmospheric and
  solar mass squared splittings, respectively;
  see~\cite{Beringer:1900zz}. Once $\theta_{24}$ becomes $O(1)$ a
  three-flavor analysis from which these parameters are inferred from
  is not strictly self-consistent.}
Hence, the new interaction $G_B$ (see below) is protected for $G_B/G_F
\lesssim 10^4$ from laboratory probes on any terrestrial baselines;
matter effects will not spoil the decoupling of the fourth neutrino
state in the $\Delta m \to 0$ limit.

To illustrate the convenience of the
parameterization~(\ref{eq:maximsU}) consider the opposite definition,
$\tilde U = R(\tilde\theta_{24}) V_3 $ together with the appearance
probability of $\nu_b$ from a muon neutrino beam when atmospheric
splitting dominates,
\begin{align*}
\sin^2 2\tilde\theta_{b\mu}^{\rm eff} & = \cos ^4\theta _{13} \sin
^4\theta _{23} \sin ^2 2\tilde\theta _{24} \quad (\tilde\theta
_{14},\tilde\theta_{34} =0).
\end{align*}
Here, $\nu_b$ appearance is not protected by a mass-splitting unless a
special combination with additional new angles $
\tilde\theta_{14},\tilde\theta_{34} \neq 0 $ is chosen, obscuring the
decoupling in the $\Delta m \to 0$ limit.

While the idea explored in this paper uses mass splittings that are
comparable to the earth-sun distance, one could also entertain a more
radical use of the ``neutrino oscillation portal'', when {\em
  e.g.}~neutrinos of astrophysical origin oscillate into the more
interacting counterparts~\cite{Joseftalk}, leading to interesting
phenomenological consequences that wait to be explored.  Also, our
framework is, of course, broader than the $4$-neutrino states. One
could also contemplate modifications to this model by the presence of
several new neutrino states, where only part of them have enhanced
interactions. This way one may also accommodate an explanation of the
short base line anomalies.

\section{Baryonic neutrinos}
\label{sec:nub}

Here we consider gauged baryon number and a new left-handed neutrino
which is charged under U(1)$_B$.  The covariant derivatives of the
baryonic neutrino $\nu_{b}$ and of the quark fields $q$ are,
\begin{align}
 D_{\mu} \nub & =  (\partial_{\mu} + i q_{\nu} g_B V_{\mu})  \nub , \\
 D_{\mu} q & =  \left(D^{\mathrm{SM}}_{\mu} + i q_B g_B V_{\mu}\right) q .
\end{align} 
$g_B>0$ is the U(1)$_B$ gauge coupling and $q_{\nu} = \pm 1$ and $q_B
= 1/3$ are the charges of $\nub = \frac{1}{2}(\mathbb{1}-\gamma^5)
\nu_b$ and quarks $q$, respectively; $|q_{\nu}|\neq 1$ can be
reabsorbed into the value of $g_B$. For $q_{\nu}=1$ and
$\theta_b<\pi/4$, the MSW condition can only be met by
$\bar\nu_b$~\cite{Pospelov:2011ha,Harnik:2012ni,Pospelov:2012gm}. $V_{\mu}$
mediates the interaction, which carries mass $m_V$ from the Higgsing
of U(1)$_B$ or from a Stueckelberg mechanism. Notice that the gauge
anomaly in the model can be cancelled due to new fermionic states at
the electroweak scale.

It is useful to measure the new interaction in units of Fermi's
constant. We define, 
\begin{align}
\label{eq:GB}
  G_B \equiv \frac{q_{\nu} g_B^2}{m_V^2} \approx 10^5 G_F \times
  q_{\nu}g_B^2 \left( \frac{1\,\GeV}{m_V} \right)^2 .
\end{align}
Invariably, a sizable enhancement, $G_B/G_F\gg 1$, 
 requires $m_V$ to be well below the weak scale. 
 As it turns out, the "safest" phenomenological choice is $m_V$ in the
 $\sim O(1-100)$ MeV range, where $g_B$ can be quite small, below
 $0.01$, and indeed difficult to detect despite kinematic
 accessibility for many experiments; see,
 \textit{e.g.}~\cite{Friedland:2011za}.

In the previous
works~\cite{Pospelov:2011ha,Pospelov:2012gm} an effective enhancement
factor was defined,
\begin{align}
  \mathcal{N}_{\mathrm{eff}}^2 \equiv \frac{1}{2} \left(
    \frac{G_B}{G_F}\right)^2 \sin^2{2\theta_b} \simeq 2\theta_b^2 \left(
    \frac{G_B}{G_F} \right)^2 ,
\end{align}
such that for low-energy processes, in the limit of rapid
oscillations, one can make the identification $P_{b \alpha}G_B^2 \to
\mathcal{N}_{\mathrm{eff}}^2 G_F^2 $.  Together with the mass splitting,
the effective enhancement factor defines a simple two-parameter 
space $\{  \Delta m, ~\mathcal{N}_{\mathrm{eff}}\} $ that we are set 
to explore in connection with the direct detection experiments.

While the baryonic portal is not unique, as one could also consider
interactions with electric charge via the more familiar ``photon
kinetic mixing'' portal, phenomenologically such option seems to be
less attractive. Indeed, new neutrino states appearing through the
oscillation of solar neutrinos are also constrained by their
neutral-current type inelastic interactions with nuclei. As shown in
\cite{Pospelov:2011ha}, the baryonic current portal is by far the
least constrained choice, while the kinetic mixing portal may not be
allowed at an interesting level of
$\mathcal{N}_{\mathrm{eff}}$. Finally, $\nu_b$ is also a safe option
when it comes to cosmological
aspects~\cite{Pospelov:2011ha,Dasgupta:2013zpn}.

\section{Direct Detection}
\label{sec:dd}

The spin-independent elastic recoil cross section on nuclei
can be obtained from the usual active neutrino-nucleus coherent
scattering~\cite{PhysRevD.30.2295} using the replacement
$G_F^2(N/2)^2\to G_B^2 A^2$~\cite{Pospelov:2011ha}.
In terms of a recoil cross section,
\begin{align}
\label{eq:cs}
  \frac{d\sigma_{\mathrm{el}}}{d\Er} & = \frac{G_B^2}{2\pi} \,A^2
  m_N  F^2(|\mathbf{q}|)\left[ 1 - \frac{(E_{\mathrm{min}})^2}{\Enu^2} \right] ,
\end{align}
where $E_{\mathrm{min}} = \sqrt{\Er m_N/2}$ is the minimum energy
required to produce a recoiling nucleus of mass $m_N$ and kinetic
energy $\Er$. $A$ is the atomic number of the nucleus, and  the nuclear form
factor suppression is given by $F^2(|\mathbf{q}|)$ for scatterings
with momentum transfer~$\mathbf{q}$. We employ the Helm
parametrization~\cite{Helm:1956zz} with the nuclear skin
thickness of~0.9\,\fm.

As it turns out, \bet\ neutrinos from the sun have the best
combination of large flux and high end-point energy, $\Phi_{\bet} =
(5.69^{+0.173}_{-0.147})\times 10^6\,\cm^{-2}\,\sec^{-1},$ and $
E_{\mathrm{max}} = 16.36\,\MeV $~\cite{Bahcall:2004pz},
respectively.  The MSW solution to the solar neutrino problem operates
on the highly energetic part of the neutrino spectrum so that $\nu_e$
exit the sun mostly as $n_2$. It is precisely this part of the
spectrum that is relevant for producing nuclear recoils above
detection thresholds.
With $\theta_b = \theta_{24} \neq 0$ the appearance probability from
the solar $\nu_e$ flux in the tri-bimaximal mixing approximation is given
by~\cite{Pospelov:2011ha},
\begin{align}
\label{eq:Peb}
  P_{be}(L,\Enu) \simeq \sin^2(2\theta_b) \sin^2{\left[ \frac{\Delta m^2
        L(t)}{4 \Enu} \right]} . 
\end{align}

The theoretical recoil spectrum arising form a solar \nub\ flux is
given by a convolution of the recoil cross section~(\ref{eq:cs}) with
the neutrino differential flux $df/d\Enu$ weighted by $P_{be}$,
\begin{align}
\label{eq:rate}
\frac{dR}{dE_R} = N_T \left[ \frac{L_0}{L(t)} \right]^2 \Phi_{\bet}
\int^{\mathrm{E_{\mathrm{max}}}}_{\mathrm{E_{\mathrm{min}}}}
dE_{\nu}\, P_{be}(t,E_{\nu})\frac{df_i}{dE_{\nu}}
\frac{d\sigma_{\mathrm{el}}}{dE_R} .
\end{align}
We have included an overall flux modulation factor $[L_0/L(t)]^2$ due
to the earth's eccentric orbit (ellipticity $\epsilon = 0.0167$)
around the sun with a maximum at perihelion on $\sim$Jan~2 and $L_0 =
1~{\rm AU}$. For $df/dE$ we take the \bet\ flux
from~\cite{PhysRevC.54.411,PhysRevC.56.3391}; \mbox{$\int dE\,
  df_i/dE=1$}. $N_T$ denotes the number of target nuclei per unit
detector mass. 

Equipped with the necessary theoretical quantities we now turn to
those experiments which have either released new data or received
significant upgrades in their understanding of detector and signal;
for details on other searches and how their results are translated to
the $\nu_b$ model we refer the reader to our previous
work~\cite{Pospelov:2012gm}.

\subsection{Xenon experiments}
\label{sec:Xenonexps}

Rare underground event searches based on liquid/gaseous xenon
two-phase experiments have advanced to the workhorses of DM direct
detection. The advantage lies in scale-ability and the fiducialization
of detector mass. 

Discrimination between nuclear and electron recoils is achieved by
measuring prompt (S1) and delayed (S2) scintillation light. The latter
is due to drifting ionized electrons into the gaseous phase.  Nuclear
recoil energies are obtained from
\begin{align}
\label{eq:LeffEr}
{\rm S}1 =  \Er \mathcal{L}_{\mathrm{eff}}(\Er) L_y\frac{S_n}{S_e}, 
\qquad 
{\rm S}2 = \Er Q_y(\Er) Y ,
\end{align}
measured in units of photo-electrons (PEs).
Only a small fraction of the deposited recoil energy is emitted in
form of prompt scintillation light S1. It determines the
discrimination threshold and relies upon the scintillation efficiency
$\mathcal{L}_{\mathrm{eff}}$ for nuclear recoils relative to electron
recoils; $L_y $ is the light yield in PEs/\keVee\ of a
$\gamma$-calibration source and $S_e$ and $S_n$ are experiment-specific
quenching factors for scintillation light due to electron and nuclear
recoils, respectively.
$Q_y$ is the ionization yield per keV nuclear recoil and $Y$ is the
measured number of PEs produced per ionized electron which is an
amplification factor.

\subsubsection{LUX}
\label{sec:LUX}

The Large Underground Xenon (LUX) experiment located at the Sanford
Underground Research Facility has very recently published its first DM
data from 85.3 live days with a fiducial target mass of
118~kg~\cite{Akerib:2013tjd}. A total of 160 events were observed in
the S1 acceptance region $2-30$~PE, consistent with electron-recoil
(ER) background from radioactive contamination.  LUX is now the
leading experiment in the search for DM spin-independent scatterings,
excluding a cross section of $8\times 10^{-46}\,\cm^2$ at DM mass
$33\,\GeV$.

Limits on the $\nu_b$ signal are derived as follows. We first compute the
expected S1 signal from $\nu_b$ using the absolute light yield
estimation (the quantity $\mathcal{L}_{\mathrm{eff}}L_y S_n/S_e$)
of~\cite{Szydagis:2013sih, LUXtalk}. Conservatively, a hard cut is
introduced at $E_R = 3\,\keV$. Next we account for the Poisson nature
of the process and retain events in the S1 acceptance window
$2-30$~PE. The signal is then binned in accordance with supplementary
Fig.~8 which reports observed and ER-background rates as a function of
S1. This allows us to account for expected background
$\nu_{\mathrm{bg}}$ and we use the first five bins $N_{\rm bins}= 5$
with $S1\lesssim 26~$PE where $\nu_b$-signal $ \nu_{\mathrm{s}}$ is
present; $\nu = \nu_{\mathrm{s}} + \nu_{\mathrm{bg}}$. A value of
$\Neff$ is excluded in each bin at a level $1-\alpha_{\mathrm{bin}}$
if the probability to see as few as $n_{\mathrm{obs}}$
Poisson-distributed observed events is $\alpha_{\mathrm{bin}}=
\sum_{n=0}^{n_{\mathrm{obs}}} \nu^n \exp(-\nu)/n!$. The overall
exclusion limit is obtained by including a statistical penalty for
observing more than one bin, $1-\alpha =
(1-\alpha_{\mathrm{bin}})^{N_{\mathrm{bin}}}$.

The resulting constraint is shown in Fig.~\ref{fit} by the thick solid
line. Its modulation in $\Neff$ as a function of $\Delta m^2$ is a
generic feature of all limits. For $\Enu \sim 10\,\MeV$ and $\Delta
m^{2} = O(10^{-10}\,\eV^2)$ the oscillation length $L_{\rm osc }= 4\pi
\Enu /\Delta m^2$ attains the value of the earth-sun
distance,~1~AU. For $\Delta m^2 > 10^{-9} \,\eV^2 $ the oscillation
becomes rapid so that $\sin^2 [\Delta m^2 L/(4 \Enu)] \to 1/2$ and all
limits turn into horizontal lines, independent of $\Delta m^2$.

\subsubsection{XENON10, XENON100}
\label{sec:Xenon10-100}

Before LUX, XENON100~\cite{Aprile:2010bt,Aprile:2011hi} set the most
stringent limits on DM scattering via spin-independent scattering.
However, for the $\nu_b$ model, the low-energy ionization-only
analysis by XENON10~\cite{Angle:2011th} with a pre-cut exposure of
15~kg~days turned out to be the most constraining experiment. This is
due to the steeply falling recoil spectrum a practically massless
particle of MeV energies can induce.

Here we improve on our previous analysis in two significant
ways. First, for XENON100 the final data set with an exposure of
$225\,\days\times 34\,\kg~$ is published~\cite{Aprile:2012nq}. Since
only two events consistent with background are observed, previous
sensitivity to $\nu_b$ from $100$~live days is approximately doubled.
Second, the XENON10 ionization-only low-threshold study underwent some
revision by the collaboration. Importantly, systematic errors of $Q_y$ have
been assessed in~\cite{Aprile:2013teh}. We can now derive a limit from
XENON10 that is better corroborated. In our previous work, the XENON10
limit was most uncertain of all, due to the delicate sensitivity on
$Q_y$ below $\Er<10\,\keV$ for which $Q_y$ is  poorly known.

Our derivation of XENON100 limits is based on a 99.75\% ER rejection
cut and is in methodology similar to our previous
analysis~\cite{Pospelov:2012gm}. Therefore, in contrast to LUX, we are
not required to model ER background. Two events remain in the final
data-set; they are consistent with a background expectation of $(1\pm
0.2)$~\cite{Aprile:2012nq}. 
The S1 signal is computed using the parameterization of
$\mathcal{L}_{\rm eff}$ given in~\cite{Aprile:2011hi} which is based
on the measurements of~\cite{Plante:2011hw}. In accordance with our
previous analysis~\cite{Pospelov:2012gm}, we extrapolate
$\mathcal{L}_{\mathrm{eff}}$ linearly to zero from $3\,\keV$ to
$2\,\keV$.%
\footnote{If we adopt instead the more conservative approach and
  refrain from modeling nuclear recoils below $3\,\keV$ the resulting
  limit becomes instead weaker with respect
  to~\cite{Pospelov:2012gm}. This demonstrates the great sensitivity
  on $\mathcal{L}_{\rm eff}$ given the soft recoil spectrum $\nu_b$
  induces for xenon.}
We take into account Poisson fluctuations in S1 and set the analysis
threshold to 3~PE in accordance with~\cite{Aprile:2012nq}.
New cut acceptances are found in Fig.~1 of the latter paper and $L_y =
(2.28 \pm 0.04) $ is a value updated in the same reference; $S_e =
0.58$, $S_n=0.95$, and $Y = 19.5$~PE.
The two observed events had $S1=3.3$~PE and $3.8$~PE. Using Yellin's
maximum gap method~\cite{Yellin:2002xd} we set an upper limit on the
$\nu_b$ signal strength.  The resulting constraint is shown in
Fig.~\ref{fit}.

We now turn to XENON10. The S1 signal is discarded so that with the
S2-only analysis a lower threshold of $\Er = \mathcal{O}(1\,\keV)$ can
be reached---at the expense of additional background. 
In an early revision by the collaboration, a position cut was
discarded so that the number of events in the energy window between
$\sim$1.4-10\,\keV\ increases from 7 to 23 with effective exposure
more than doubled, from $6.2\,\kg\,\days$ to $13.8\,\kg\,\days$;
Fig.~2 in~\cite{Angle:2011th}.
In addition, the new analysis in~\cite{Aprile:2013teh} quantifies
potential systematic errors in $Q_y$, and hence in the calibration of
nuclear recoil energy, for $\Er \gtrsim 3\,\keV$. Here we derive the
limit based on $Q_y$ shown in Fig.~3 of~\cite{Aprile:2013teh}. For
this, we group the observed events into three bins of 5--32 ionized
electrons, $\{ n_e^{\rm min}, n_e^{\rm max}, N_{\rm obs} \}$, namely,
$\{5,8,2\}$, $\{8,16,5\}$, and $\{16,32,9\}$ with a total number of
16~events observed. The computation of S2 then proceeds similarly to
S1 above where we account for Poisson fluctuations in the number of
ionized electrons.
A final extrapolation of $Q_y(\Er)$ to $\Er(5\,e^-)$ is necessary.  We
assume a flat trend. Varying $Q_y$ within its $\pm 1\sigma$ error band
with the same extrapolation changes the limit only very mildly within
$\Delta \Neff < 10.$ The exclusion limit based on the binned data set
is derived identically as in the LUX case.

\subsection{CDMS-Si}
\label{sec:cdms}

A blind analysis by the CDMS-II collaboration of 140.2 kg-days of data
collected with their silicon detectors revealed three DM-candidate
events with a total expected background of
0.7~\cite{Agnese:2013rvf}. When including the spectral information of
the observed events, the background-only hypothesis has a 0.19\%
probability when compared to one which includes the signal of a WIMP
with spin-independent cross section of $1.9\times 10^{−41}\,\cm^2$ and
mass $8.6\,\GeV$. This interpretation of CDMS-Si result is intriguing but is now severely
challenged by the null-observation of LUX, where more than $10^3$
events were expected at the CDMS best fit point~\cite{LUXtalk}. Here
we offer an alternative explanation of the observed excess that at
this point stands unchallenged by LUX.

\begin{figure}
\centering
\includegraphics[width=\columnwidth]{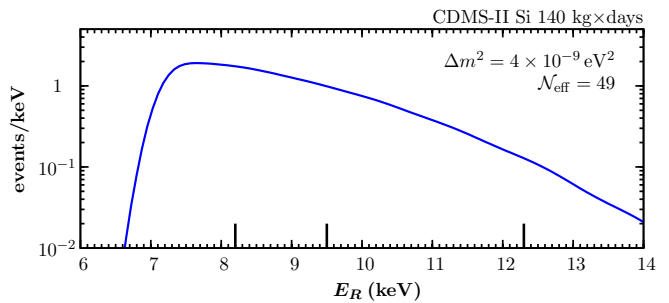} 
\caption{\small Fit of the three silicon events in CDMS-II; the recoil
  locations of the events are indicated by the vertical bars at the
  bottom of the figure.}
\label{cogent}
\end{figure}

We fit the tree events using $1\,\keV$ bins starting from $7~\keV$
which corresponds to the lowest threshold of all detectors untill
$15~\keV$ above which no signal from $\nu_b$ is induced. The overall
efficiency is taken from Fig.~1 of~\cite{Agnese:2013rvf}. The best fit
is inferred from minimizing the Poisson log-likelihood ratio
\begin{align}
\label{eq:poisson-L}
  \chi^2 = 2 \sum_{{\rm bins\ } i} \left[ y_i - n_i + n_i \ln{\left(
        \frac{n_i}{y_i} \right)} \right] ,
\end{align} 
where the last term absent for zero observed events, $n_i = 0 $, $y_i$
is expected number of events.
The minimum in $\chi^2$ is attained for,
\begin{align}
\label{eq:cdms-bestfit}
{\rm CDMS-Si:}\,\,\Delta m^2 = 4\times 10^{-9}\,\eV^2,\,\,\Neff = 49
\end{align}
with $\chi^2/n_d = 7/6$. The favored region is shown in
Fig.~\ref{fit}. As one can see, there is no special preference for
small $\Delta m^2$, and the allowed band can be continued to fairly
large values to be cut off only by a modification of neutrino
oscillations in short- and long-baseline type of experiments.

\subsection{CoGeNT}
\label{sec:cogent}

\begin{figure}
\centering
\includegraphics[width=\columnwidth]{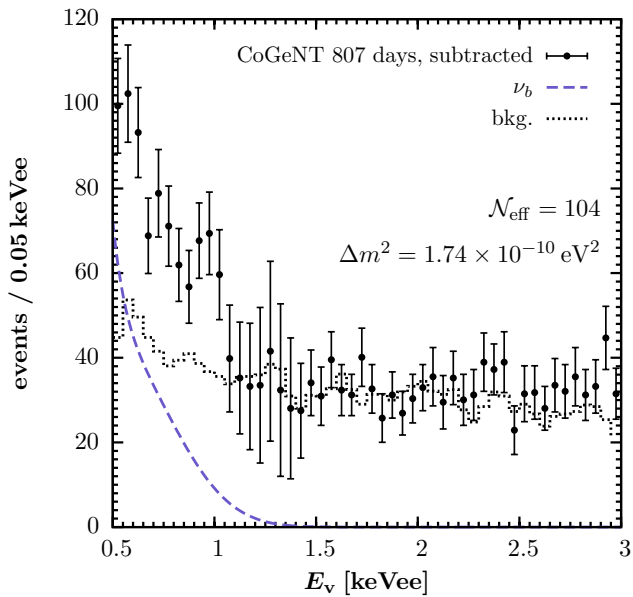} 
\caption{\small Fit to the CoGeNT 807 live-day spectrum. The data
  points show the signal after subtraction of cosmogenic radioactive
  background. The dotted line represents the best understanding of
  backgrounds according to~\cite{Aalseth:2012if}; when the dashed line
  from $\nu_b$ is added is a good fit to the black data points is
  obtained.}
\label{cogent}
\end{figure}

The CoGeNT
experiment~\cite{Aalseth:2010vx,Aalseth:2011wp,Aalseth:2012if} employs
a p-type point-contact germanium crystal ($0.44\,\kg$). It has the
advantage of a very low energy threshold $0.5\, \keVee$ electron
recoil equivalent so that despite a moderately heavy target nucleus
$(A=76)$, good sensitivity to light DM in the $10\,\GeV$-ballpark or
to $\nu_b$ is expected. Indeed, the collaboration reports an
unexplained quasi-exponential rise at lowest energies.%
\footnote{In addition to the signal-rise below 1\,\keVee\, the data
  also appears to be annually modulated in the 0.5--3.2\,\keVee\
  bracket which we do not address here.}
The origin of it is unknown and has lead to the speculation of a DM
signal with favored mass in the $\sim 8-10\,\GeV$. This hypothesis is
now seriously challenged, if not completely excluded by LUX.

The drawback of the experiment is that it registers ionization from
both, nuclear recoils and electromagnetic background without being
able to discriminate between the two.  Already at the time of our
previous analysis~\cite{Pospelov:2012gm}, it was clear that incomplete
charge collection on the surface induces a potential background that
mimics a signal in $\nu_b$. Hence in~\cite{Pospelov:2012gm} reliable
ROIs could not be provided.
Instead, an envelope that limits $\Neff \lesssim 200$ was established.
It indicated the region of ``compatibility'' with the other
anomalies. Larger values of \Neff\ lead to signal strengths in excess
of the observed data.

\begin{figure*}
\includegraphics[width=0.85\textwidth]{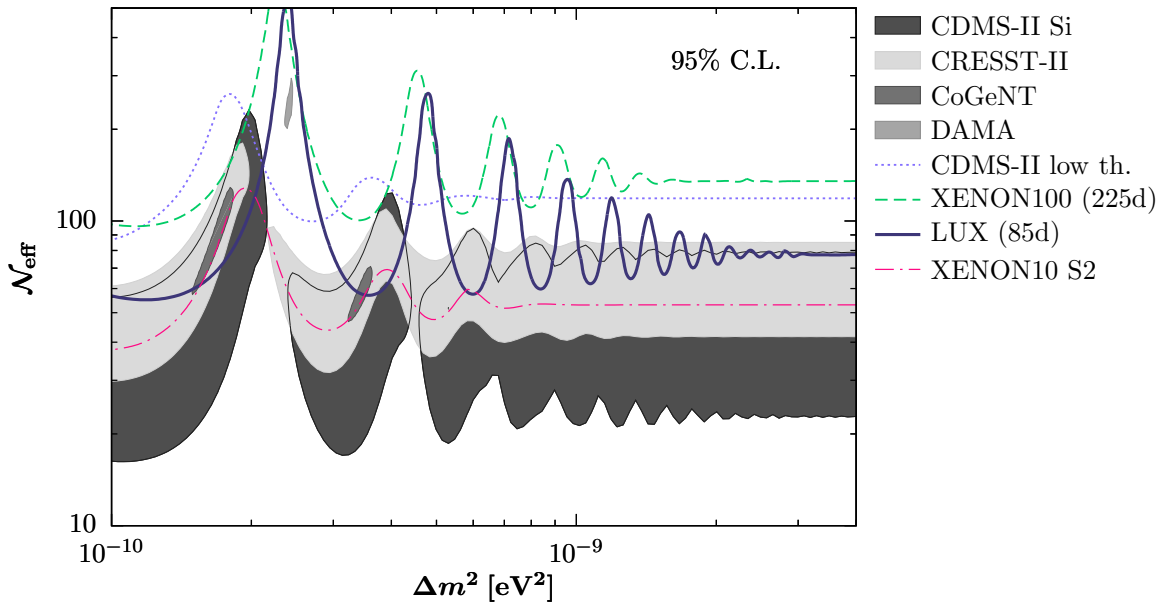} 
\caption{\small Summary plot of direct detection favored regions and
  constraints in the parameters $\Delta m^2$ and \Neff\ at 95\%
  confidence. \textit{Favored regions:} the broad (dark) light shaded
  gray band shows the CDMS-Si (CRESST-II) regions. The two small shaded
  islands are the regions in which the CoGeNT excess is explained.  In
  the medium gray shaded island DAMA's modulation amplitude is fitted;
  the phase, however, remains significantly discrepant by about one
  month.  \textit{Constraints:} \Neff\ values above the various
  lines are excluded with 95\% confidence.  From top to bottom at 
  $\Delta m^2 > 10^{-9}\,\eV^2$ the respective constraints are from XENON100, from 
  CDMS-II low-threshold data, from LUX, and from XENON10.
}
\label{fit}
\end{figure*}

In~\cite{Aalseth:2012if} the collaboration provides a detailed
discussion of backgrounds. Therefore, we are now able 1) to infer ROIs
that are based on a better, quantitative understanding of backgrounds
and 2) to use a data set which almost has twice the exposure of that
of our previous analysis (807 vs.~442~kg~days). The data together with
the background model are taken from Fig.~23 of~\cite{Aalseth:2012if}
and are respectively reproduced by the points and dotted line in
Fig.~\ref{cogent}.  The spectrum is corrected for efficiencies and cut
acceptance; cosmogenically induced radioactive backgrounds have been
subtracted.
We use Lindhard theory to convert nuclear recoil energies to
ionization signal that is calibrated with gamma radiation,
$\Ev(\keVee) = Q \times \Er(\keV)^{1.1204}$. The quenching factor is
$Q = 0.19935$ and the detector resolution is $\sigma^2 = (69.4\,\eV)^2
+ 0.858\,\eV\times\Ev(\eV)$~\cite{Aalseth:2008rx}.

The maximum recoil energy of \bet\ neutrinos on germanium is $\Er
\leq 7\,\keV$ which corresponds to an electron recoil equivalent of
$E_{\rm v} \leq 1.76\,\keVee$. This coincides with the region in which
the exponential rise is present and we fit the first 25 data points in
that energy interval. With $95\%$ confidence we find two ROIs. The
best fit is attained for
\begin{align}
\label{eq:cogent-bestfit}
{\rm CoGeNT:}\,\,\Delta m^2 = 1.74\times 10^{-10}\,\eV^2,\,\,\Neff = 104
\end{align}
and is shown by the solid (blue) line in Fig.~\ref{cogent}.
The model provides an excellent description of the data, $\chi^2 /n_d
= 18/23 $. We observe a significant shift of the best-fit ROI with
respect to our previous analysis~\cite{Pospelov:2012gm}. In the
latter, given the absence of quantified backgrounds, the signal-only
interpretation favored a best-fit ROI centered around $\Neff =
228$. With the better understanding of backgrounds, the CoGeNT
explanation now overlaps with the regions inferred for CDMS-II silicon
and CRESST-II.

As one can see, only small fraction of the parameter space of the
$\nu_b$ model can explain the CoGeNT excess. In particular, it appears
that large values of $\Delta m^2$ are incompatible with current
results.  This stems from too soft of a recoil for most of the
oscillation patterns except the one that gives relative enhancement to
the most energetic part of $^8$B spectrum, yielding events with
$E_{\rm v} \sim 1\,\keVee$. It must be said that if the current model
of background introduces some spectral distortion, and the excess is
actually {\em softer} than appears in Fig.~\ref{cogent}, then the
available parameter space can be significantly enlarged, with $\Delta
m^2$ in excess of $10^{-9}\,\eV^2$ also becoming open for speculative
interpretations.

\section{Similarity with light-DM}
\label{sec:simil-with-light}

\begin{figure}
\includegraphics[width=\columnwidth]{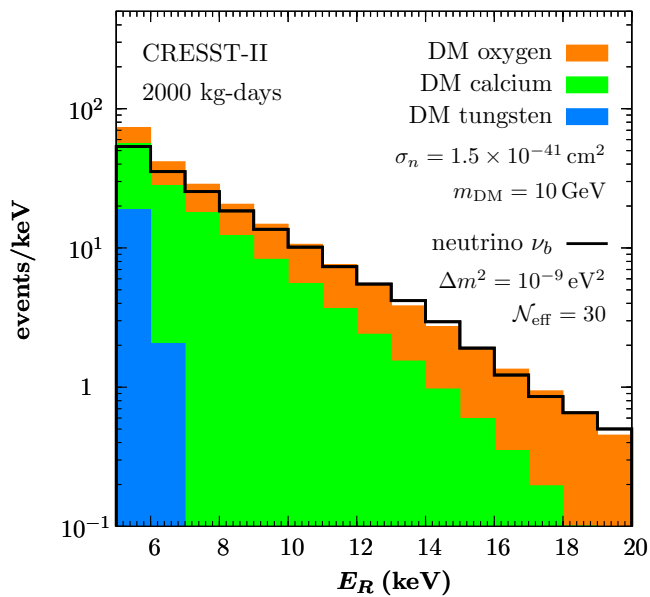}
\caption{\small Comparison of theoretical recoil spectra from $\nu_b$
  with $\Neff = 30 $, $\Delta m^2 = 10^{-9}\,\eV^2$ and from DM with
  $m_{\rm DM} = 10\,\GeV$, $\sigma_{n} = 1.5\times 10^{-41}\,\cm^2$,
  and $v_{\rm esc} = 544\,\km/\sec$ in a future CRESST-II data
  set. Above $E_R=10\,\keV$ (a realistic threshold of the experiment)
  all $\nu_b$ recoils are on O.}
\label{cresst}
\end{figure}

It is known that light DM and $\bet$ neutrinos from the sun can induce
recoil spectra in direct detection experiments that are similar in
morphology. Hence a positive signal of either origin may be confused
with one of the other. The ballpark of DM masses $m_{\rm DM}$ where
this is the case can be found by comparing the DM maximum recoil
energy $E_R^{\rm max} = 2 m_{\rm DM}^2 v^2/m_N$ $(m_{\rm DM}\ll m_N)$
with the one from neutrinos, $E_R^{\rm max} = 2\Enu^2/m_N$. Hence,
$m_{\rm DM}\approx \Enu/v = O(8)\,\GeV$ for typical values of $\Enu$
and the DM-target relative velocity $v$.
This similarity of signal is illustrated in Fig.~\ref{cresst} for a
projected CRESST-II data set on a CaWO$_4$ target.

In the following the two alternative hypotheses for a potential signal
in future direct detection experiments are explored.  The $\nu_b$
model is very predictive once the oscillation length becomes smaller
than the earth-sun distance. The only free parameter is then $\Neff$
and it renormalizes the signal strength. A straightforward approach is
to generate recoil spectra from $\nu_b$ (the underlying true
hypothesis) and ask the question: How easily are the observed data
confused with a detection of light DM (the alternate hypothesis)?

To quantitatively answer this question we produce mock sets of data
for various target nuclei for the exemplary values $\Delta m^2 =
10^{-9}\,\eV^2$ and $\Neff = 30$. 
In anticipation of a future superCDMS setup where some fraction of
detectors are made from silicon, we generate spectra for Ge and Si
with nuclear recoil energies in the interval $\Er =5-8\,\keV$ and
bin-size of $0.5\,\keV$ and respective exposures of 20~kg-yr and
100~kg-yr. With the chosen parameters, the total number of events in
Ge (Si) is 45 (1268). For germanium 90\% of events are located in the
first bin so that in contrast to silicon, rather limited spectral
information is available.

In addition to superCDMS, we also provide projections for the
CRESST-II upgrade that is---at the time of writing---taking data. New
detector designs have been developed and show promising results in the
elimination of previously encountered Pb and $\alpha$
backgrounds~\cite{CRESSTtalk}. The projected joint exposure of their
CaWO$_{4}$ detectors of a 2~year run is 2000~kg-days with the
potential to conclusively test its own anomaly.
We generate a mock spectrum from $\nu_b$ with same neutrino parameters
as above but with a higher threshold of $10\,\keV$, which more closely
resembles the experimental reality. The spectrum in oxygen extends to
higher recoil energies with a total number of 36 expected events in 10
bins in the energy interval $10-20\,\keV$, all of which are oxygen
recoils.

\begin{figure}
\includegraphics[width=\columnwidth]{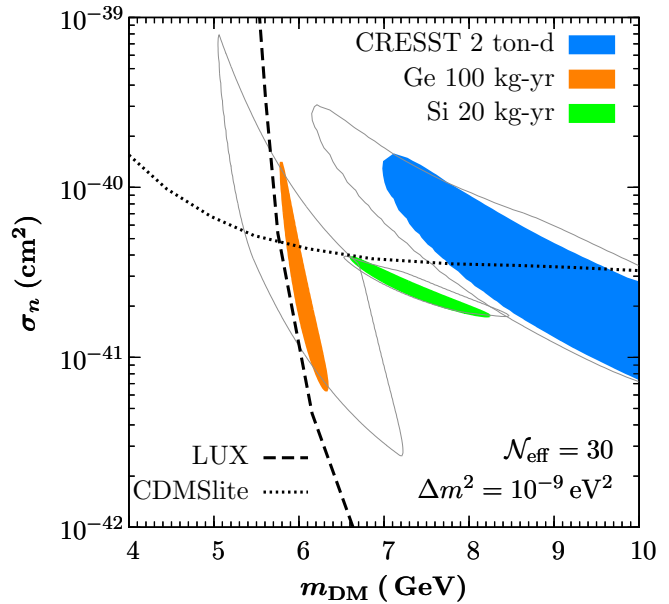}
\caption{\small Dark matter interpretation of a positive $\nu_b$
  signal with $\Neff=30$ and $\Delta m^2 = 10^{-9}\,\eV^2$ in a
  future (idealized) superCDMS-type direct detection experiment
  equipped with Ge and Si targets as well as for the new run by the
  CRESST-II dark matter search.  The exposures are 20 (100) kg-yr for
  Si (Ge) and 2000\,kg-days for CaWO$_4$. Inside the shaded regions,
  the DM hypothesis is (falsely) accepted with p-value $p>5\%$ by the
  goodness-of-fit criterion. From the disjointness of the regions DM
  and $\nu_b$ can be discriminated which stresses the importance of
  employing multiple targets. Inside the parameter regions, enclosed
  by the thin gray lines, the DM escape velocity is allowed to vary
  between $450 \leq v_{\rm esc}(\km/\sec)\leq 650$, indicating the
  areas of parameter space in which the shaded regions can move for
  the Maxwellian halo model. The LUX and CDMSlite exclusion limits are
  as reported by the collaborations. }
\label{DM}
\end{figure}

The binned spectra are then fitted to a canonical light DM model with
spin-independent WIMP-nucleon cross section $\sigma_n$
using~(\ref{eq:poisson-L}). We choose a Maxwellian DM velocity
distribution with escape speed $v_{\rm esc }$ and most probable
velocity $v_0 = 220\,\km/\sec$.
The goodness-of-fit serves as the test statistic for the DM
hypothesis, \textit{i.e.}~DM is considered as the null hypothesis and
we (falsely) accept it for a p-value $p>5\%$.
The result of the fit is shown by the shaded regions in Fig.~\ref{DM}
for $v_{\rm esc}=544\,\km/\sec$~\cite{Smith:2006ym}. Outside of them,
the probability to confuse the $\nu_b$ signal with light DM from the
observation in either Ge, Si, or O is less than $5\%$. Also shown are
current limits on the SI cross section from LUX and
CDMSlite~\cite{2013arXiv1309.3259A}. In order to quantify the
dependence on the velocity distribution in the regions enclosed by the
thin gray lines $v_{\rm esc}$ is allowed to float in the range $450
\leq v_{\rm esc}(\km/\sec)\leq 650$; of course, the same halo
parameters will hold for Ge and Si so that no effective broadening as
indicated by the thin gray lines takes place in reality. Rather, the
thin lines indicate the ``comfort zone'' in which the ROI can move
around for the Maxwellian halo model.%
\footnote{The reason why the lines intersect slightly with their
  associated shaded regions is because in the former (latter) case the
  goodness-of-fit statistic follows a $\chi^2$ distribution with
  $N_{\rm bins}-3$ ($N_{\rm bins}-2$) degrees of freedom; the limits
  from LUX and \mbox{CDMSlite} are nominal ones---they too will be
  affected by changes in $v_{\rm esc}$.}

The first observation is that all shaded regions are disjoint.  As
alluded before, the spectrum in Ge falls steeply and has little
statistics. Therefore a light DM particle of mass $m_{\rm DM}\simeq
6\,\GeV$ is selected. Exponential sensitivity in the mass translates
into a region in WIMP-nucleon cross section that can vary by one order
in magnitude, $\sigma_n \sim 10^{-41} - 10^{-40}\,\cm^2$. With much
more statistics the ROI in Si is significantly smaller and favors
heavier DM. A more shallow signal yields a bigger allowance in $m_{\rm
  DM}\simeq 7-8\,\GeV$ but selects a more narrow range of cross
sections, $\sigma_n \sim 2-3\,\times 10^{-41}\,\cm^2$, and which
serves to normalize the rate.
Finally, recoils on O in a future CRESST-II data set contain both
features: small statistics gives a relatively accommodating region in
parameter space and a shallow spectrum that allows for a very broad
range in DM mass.
The fact that the ROI are largely disjoint stresses the importance of
employing multiple targets in direct detection. It breaks the
degeneracy between these two alternative new physics interpretations.

Forecasts can also be made for the liquid scintillator experiments
such as XENON1T or future exposures of LUX. For the xenon target, the
maximum recoil energy of \bet\ neutrinos is $\Er \leq 4.5\,\keV$ so
that the most promising avenue lies in the exploration of
ionization-only, low-threshold studies.
To a fair extent, we have covered this ground in our previous
paper~\cite{Pospelov:2012gm} where we showed the projections for
XENON100 and that can be rescaled for larger exposures.  The finding
was that better sensitivity is mostly limited by ER backgrounds. For
the XENON100 detector they are at the level of
$10^{-2}\,\cpd/\kg/\keV$~\cite{Aprile:2011vb}; the LUX measured ER
rate in the fiducial volume in the keV-regime is now $3\times
10^{-3}\,\cpd/\kg/\keV$~\cite{Akerib:2013tjd}.
Therefore, we do not expect immediate drastic improvements so that we
refrain from repeating a similar analysis here but instead await new
data from the experiments.

The analysis presented here can be further extended and improved:
First, in CRESST-II, the phonon measurement provides a nuclear recoil
energy scale so that the quenching of the scintillation light signal
allows for a (limited) discrimination between recoils on Ca, O, or
W. The $\nu_b$ model does not induce W-recoils (and no Ca recoils for
$\Er >10\,\keV$, as can be seen from Fig.~\ref{cresst}) so that
further differential information is available that will help to
scrutinize the model from light WIMPs. The discrimination of nuclear
bands works better of high recoil energies and it becomes a
quantitative question how well this can be done in practice.
Second, the generation of the $\nu_b$ signal was done under idealized
circumstances. Poisson fluctuations in the generation of the mock
spectrum have not been taken into account. This can be achieved in
Monte-Carlo-type analyses in which a large sample of initial spectra
is generated. Our conclusions will not be affected by this. The ROI in
Fig.~\ref{DM} may broaden somewhat but the goodness-of-fit test
already ``knows'' about the statistical error bar so that all
qualitative features will remain. More detailed explorations of this
sort and the comparison with other halo models are left for future
work.

\section{Conclusions}
\label{sec:conclusions}

In this work we have re-visited the model of a new neutrino state that
has no charged current interaction with matter, but has a new
neutral-current-type force with baryonic charge. If the strength of
the interaction is chosen to be two orders of magnitude above the
strength of the SM weak force, an important experimental probe of this
model is the coherent scattering of neutrinos on nuclei.  When the
mass splitting is small, {\em e.g.} $\Delta m^2 \ll \Delta
m_{\odot}^2$, no significant flux of new neutrinos must be produced in
the terrestrial neutrino beam experiments, and consequently the
enhanced interaction of new neutrinos is not as constrained as it may
naively seem.  However, the mixing of $n_4$ with the SM massive state
$n_2$ can easily lead to a finite oscillation probability for the
solar neutrino flux, with observable consequences for the rare
underground event searches, and especially for the Dark Matter direct
detection experiments.

A region of the WIMP parameter space with mass range of a few GeV and
cross section of few$\times 10^{-41}$ cm$^2$ per nucleon has attracted
enormous amount of attention in recent years, due to a number of
experimental results that can be interpreted as positive evidence for
the recoil signal.  However, following the release of the first LUX
results~\cite{Akerib:2013tjd}, the explanation of direct detection
anomalies via light DM models are now thoroughly disfavored. In
contrast, we find that the elastic scattering of $\nu_b$ is still an
attractive candidate for explaining most of the anomalies, capable of
a simultaneous fit to the CDMS-II silicon~\cite{Agnese:2013rvf},
CoGeNT~\cite{Aalseth:2012if}, and CRESST-II~\cite{Angloher:2011uu}
excess of events. The conclusion about the viability of the model is
based on our computation of the LUX limit on $\nu_b$, and the update
of limits from XENON100 using the final data set~\cite{Aprile:2012nq},
as well as from XENON10 with a better understanding of systematic
uncertainties in the charge yield from nuclear
recoils~\cite{Aprile:2013teh}.

Since the model of $\nu_b$ is very predictive in the most relevant
region of parameter space, we expect that in the coming years it will
be decisively tested. 
Irrespective of current anomalies, our investigation offers a future
path for distinguishing between light-DM and neutrino recoil signals.
Indeed, the ongoing low-background run of CRESST-II~\cite{CRESSTtalk},
as well as the planned expansion of the CDMS program
(superCDMS~\cite{Akerib:2006rr}) will have the potential to do so, and
in a way that is free from the uncertainties in the low-energy recoil
response, liquid xenon experiments are prone to.

Finally, while the $\nu_b$ model, as shown in this paper, is getting
constrained at an interesting level by DM direct detection
experiments, there are complementary ways of exploring the same model.
In proton-on-target type of experiments such as MiniBooNE
\cite{Dharmapalan:2012xp}, the pair-production of $\nu_b\bar \nu_b$
can be studied.
The $\nu_b$ appearance in the solar neutrino flux can also be directly
constrained at a very interesting level by the search of carbon
excitation lines in Borexino data~\cite{Bellini:2008mr}.

\begin{acknowledgments}
  JP would like to thank L.~Dai, M.~Kamionkowski, and I.~Shoemaker for
  discussions on various aspects of the baryonic neutrino model.
\end{acknowledgments}

\bibliography{biblio}

\end{document}